\documentclass[11pt]{article}
\usepackage{graphicx}
\begin{document}
%

\begin{center}
{\large \bf An Anatomy of Neutrino Oscillations}

\vskip.5cm

W-Y. Pauchy Hwang\footnote{Correspondence Author;
 Email: wyhwang@phys.ntu.edu.tw; another new version of arXiv:1209.5488v1
 (hep-ph) 25 Sep 2012} \\
{\em Asia Pacific Organization for Cosmology and Particle Astrophysics, \\
Institute of Astrophysics, Center for Theoretical Sciences,\\
and Department of Physics, National Taiwan University,
     Taipei 106, Taiwan}
\vskip.2cm


{\small(September 24, 2012; Revised: August 12, 2013; November 3, 2014;
November 3, 2015)}
\end{center}

\begin{abstract}
We ponder about neutrino oscillations; a particle without a clear
identity, a neutrino of a given flavor in fact does not satisfy the
Dirac equation (which is used to define the mass eigen-states and
mass eigen-values). This alters the basic treatment of neutrino
oscillations, in that the Dirac spinors are defined as the mass
eigen-states while the flavor states can only be given as
linear combinations of the mass eigen-states (of Dirac equations).
Even though neutrino masses are tiny, the impacts of several
neutrino masses in a single reaction or a single decay, such as
possible violation of the energy-momentum conservation, should
not be overlooked.

Among those sources of oscillating neutrinos, we point out that
the ultra high energy cosmic rays (UHECR's) such as the proton of
energy $10^{18}\, eV$ or higher, used to think of being rather
stable, can capture, in the matter media, an electron to convert
into an electron-like neutrino and a spectator neutron. This
would be the most important neutrino source of the UHECR origin.

\bigskip

{\parindent=0pt PACS Indices: 12.60.-i (Models beyond the standard
model); 12.10.-g (Unified field theories and models).}
\end{abstract}

\bigskip

\section{Prelude}

Nowadays the status has evolved to the stage such
that experimental detailed studies on how neutrinos
oscillate, among themselves or into others, and on
the rates of neutrino oscillations, are called for.
On the viewpoint of theoretical physics, it poses
us very fundamental issues, from the fact that the 
particle so elementary could change its own identity.
  
Thus, we may start up with one major caution as follows:
If we look at the momentum-space Dirac spinors,
$u(p,s)$ or $v(p,s)$, they are the mass eigen-states,
$(i \gamma\cdot p + m) u(p, s) =0$ and
$(i \gamma \cdot p - m) v(p,s) =0$. Up to this point,
there is no single quantity called "the flavor eigen-state"
except the linear combinations such as $U_{\nu_e i} u^i(p,s)$
with $u^i(p,s)$ the mass eigen-states. Thus, we have to keep
in mind the notations that are meaningful, and to avoid those
which might be meaningless. By "eigen-states", we mean that the
function(s) satisfies some differential equation, in the
sense of quantum mechanics. By this token, there is no
"neutrino flavor eigen-state" since, unlike the mass
eigen-states, we don't know how to write down a neutrino
flavor eigen-state together with the eigen-value. Of
course, we may generalize the notion slightly,
such as the parity operator and its eigen-states
and the discrete eigen-values. Without this
generalization, our criticism stands.

Or, the Dirac equations are used to define the energy
or mass eigen-states; what do we mean "flavor eigen-states"?
In writing down the S-matrix elements or the transition
amplitudes, for example, we have to write $u_{\nu_\mu}$ as
$U_{\mu i}u^i(p,s)$ with the Dirac spinor $u^i(p,s)$.
Here the word "eigen" loses the meaning of eigen, in the
sense of quantum mechanics.

On the theoretical side, the simplest way is to recognize
the existence of the "family", since neutrino oscillations
change neutrinos in one flavor (or, generation) into the
neutrinos in the other. Thus, the neutrino mass term
assumes the form \cite{HwangYan},
\begin{equation}
i {h\over 2} {\bar\Psi}_L(3,2) \times \Psi_R(3,1) \cdot \Phi(3,2)
+ h.c.,
\end{equation}
where $\Psi_L(3,2)$ is the left-handed $SU_f(3)$-triplet and
$SU_L(2)$-doublet while $\Psi_R(3,1)$ is the right-handed
neutrino $SU_f(3)$-triplet, with the first label for
$SU_f(3)$ and the second for $SU_L(2)$. The cross-dot
(curl-dot) product is somewhat new, referring to
the singlet combination of three triplets in $SU(3)$.
The Higgs field $\Phi(3,2)$ is different from the
Standard-Model (SM) Higgs, because it carries both
$SU_f(3)$-triplet charge and some $SU_L(2)$ charge
(having the charged components). Here the
dimensionless coupling $h$ (previous $\eta$), a
mass term, is determined by the size of neutrino
oscillations, such that the neutrino masses would
be in the $sub-eV$ range.

We wish to stress that the Standard Model \cite{Hwang417}
is a dimensionless theory, all couplings dimensionless
except the "ignition" term. Thus, $h$ (for neutrinos)
and $h^C$ (for charged leptons), in addition to the
$SU_f(3)$ coupling $\kappa$, govern the lepton world.

Remember that neutrino oscillations take place between
"point-like" Dirac particles, very fundamental as compared
to, e.g., oscillations in the $K^0-{\bar K}^0$ system, a
composite system. In view of the fundamentals in the group
theory, the particles would be sitting in the same multiplets
with the electrons, because of the same characteristics, and
so if the electrons are point-like Dirac particles the
neutrinos are also point-like Dirac particles. The proof
that neutrinos are point-like Dirac particles is from the
lazy man but it should be true.

Since the first version of this paper was written
(about three years ago), the author has made important
progresses (in his own judgment) and has written a
few articles related to the (new) Standard Model,
thus having implications to neutrino oscillations. In
particular, how to write a general Standard Model
was in arXiv (17 April 2013 and 25 August 2015)
\cite{Hwang417}. We also realized how to interpret
the origin of mass in a natural way \cite{Origin}.
Moreover, how to get a precise "definition" of the
Standard Model \cite{definition}. The uniqueness
of everything was stressed in the origin of fields
(point-like particles) \cite{Fields}. In view of all
these (so-called "breakthroughs"), it calls for
another updated version of this paper.

Basically, starting from three complex scaler fields, the
Standard-Model (SM) Higgs $\Phi(1,2)$, the mixed
family Higgs $\Phi(3,2)$, and the purely family Higgs
$\Phi(3,1)$ (with the first $SU_f(3)$ label and the second
$SU_L(2)$), we construct the $SU_c(3) \times SU_L(2) \times
U(1) \times SU_f(3)$ Standard Model. We realize that
$SU_f(3)$ covers only the lepton world, just like $SU_c(3)$
covering only the quark world - this results in the
separation of the lepton world from the quark world. To
study the problems of infinities, we realize that the
framework of the U-gauge together with the dimensional
regularization might provide us the first trying-out.

In our Standard Model, we {\it do not} call for particles
of new kinds, except the three Higgs fields $\Phi(1,2)$,
$\Phi(3,2)$, and $\Phi(3,1)$. Spontaneous symmetry breaking
(SSB) takes place {\it at first} at the purely family
Higgs $\Phi(3,1)$, but {\it not ignited} by the SM Higgs
$\Phi(1,2)$ \cite{Origin}. This has one important prediction,
$m_{SM} = v/2$, with $m_{SM}$ the SM Higgs mass and $v$ the VEV
in the SM sector - a result satisfied very well by the data.

The U-gauge is specified by
\begin{equation}
\Phi(1,2)= (0,{1\over \sqrt 2} (v+\eta)),\,\, \Phi^0(3,2) = {1\over \sqrt 2} (u_1+\eta'_1, u_2+
\eta'_2, u_3+\eta'_3 ),\,\, \Phi(3,1) = {1\over \sqrt 2}(w+\eta',0,0),
\end{equation}
which defines the vacuum expectation values $u_i$ in connection
with the family Higgs $\eta'_i$.

Thus, in the U-gauge, we write
\begin{equation}
i {h\over 2} \{{\bar \nu}_{\mu L}\nu_{eR} (u_1+\eta'_1) + {\bar \nu}_{eL}
\nu_{\tau R} (u_2+\eta'_2) + {\bar \nu}_{\tau L} \nu_{\mu R} (u_3 + \eta'_3)
\} + h.c.
\end{equation}
Here $u_i$ are the vacuum expectation values associated with
the family Higgs $\eta'_i(x)$. $u_i$ are given in the paper
for the origin of mass \cite{Origin}. The coupling
${\bar \nu}_{aL}\nu_{bR} u_i$ signals neutrino
oscillations.

Using the wavefunction/operator terminology, we identify neutrino
oscillations as some operator, not as part of wavefunction(s). In
the $SU_c(3) \times SU_L(2) \times U(1) \times SU_f(3)$ Minkowski
space-time \cite{definition}, there is a unique operator for
neutrino oscillations (via the last equation).

In what follows, neutrinos would be written in terms of the
mass eigen-states, and, to be careful, the neutrino flavor states,
such as $"\nu_\mu"$, would be represented as linear combinations
of the neutrino mass eigenstates, such as $U_{\mu i}u^i(k)$.

\medskip

\section{Basic Analysis of Neutrino Oscillations}

What are neutrinos? Owing to the fact that they have masses, neutrinos may
be just another point-like Dirac particles (like the electron), to be
regarded as the typical species as one kind of the building blocks
of the Standard Model \cite{Books}. In the mathematical sense, we
introduce the group(s) and the various elements - thus, we construct
the multi-plets; and the entities entering the same multi-plet should
be of the same characteristics; so, if the electrons are point-like
Dirac particles, then the accompanied neutrinos should also be point-like
Dirac particles. Even though, unlike the electrons, we cannot maneuver
the neutrinos so much these days, a direct proof that neutrinos are
point-like Dirac particles is lacking, and lacking for long time to
come.

Let's ask the question. How to write down the transition amplitude
for, e.g., ${\bar \nu}_\mu(p_1) + p(p_2) + \mu^+(p_1) + n(p'_2)$?
Without neutrino oscillations, we can easily do this in momentum
space (cf., Eqs. (6)-(7) below); with neutrino oscillations, we
cannot the Dirac spinor for the neutrino, since the Dirac spinor
is the mass eigen-state. Furthermore, is the energy-momentum
conservation at stake, since there are several assumed masses,
even very tiny, simultaneously? The conceptual problems brought
by neutrino oscillations cannot be overlooked.

In view of the basic relation Eq. (3), we see that a non-zero
mass means a non-zero vacuum expectation value (VEV) of certain form;
that is, coupling to a scalar field of certain kind. But neutrinos
oscillate - a neutrino of certain flavor can suddenly change into
that of different flavor, or ${\bar \nu_{1L}}\times \nu_{2R} \cdot
\phi$ in the mathematical form. So, neutrino oscillations only
require the off-diagonal form of the flavor states - {\it not}
the mass eigen-states at all. At this point, we have to be
satisfied with the linear combinations of the mass eigen-states,
i.e., $\sum_iU_{\mu i}u^i(p,s)$ for $"u_{\nu_\mu}"$. As we have
said, is the sacred energy-momentum conservation at stake?

There are textbooks-type reactions/decays such as
$\mu \to e + {\bar \nu}_e +\nu_\mu$, $n \to p + e^- +
{\bar \nu}_e$, etc. The oscillation interaction, Eq. (1),
enters the muon decay as well \cite{Hwang417}. To the
leading order, it does not enter the semi-leptonic beta
decays such as the neutron beta decay $n\to p+e^- +
{\bar \nu}_e$. In view of the conceptual problems
brought by neutrino oscillations, we anticipate that
there are important modifications to the
20th-century textbooks on weak interactions.

Moreover, neutrino oscillations pose another basic changes to
these textbooks, as we could write down {\it only} the
mass eigen-states while the flavor states are used in the
transient sense in the language. In a reaction, we
encounter the situation where the implementation of the
energy-momentum conservation is suffered from a final
(neutrino) state that have several masses simultaneously.

Let's come back to discuss a few issues, because of these
"peculiar" situations of neutrino oscilltions.

Using accelerators, we could produce the "high-energy" neutrino or
antineutrino beam, in the energy greater than $2\, GeV$ (i.e. the
$\tau$ mass). Considering the ${\bar\nu}_\mu - p$ scattering (the
only high-energy neutrino beam available to us, at this point), a
fraction of ${\bar\nu}_\mu$ might convert to ${\bar\nu}_\tau$,
thus making it possible to produce $\tau^+$. Similarly, for
production of $e^+$ via $\nu_\mu$.

In view of neutrino oscillations, we cannot prepare a neutrino beam
of a definitive flavor at all time - maybe it would be possible
with a neutrino beam of a definitive mass but its source is
lacking. So, we have to play with a "variable" neutrino beam in
view of neutrino oscillations.

In Figs. 1, we show the "Feynman" diagrams for the ${\bar \nu}_\mu + p$
scattering. Fig. 1(a) is the "leading" diagram for the $\mu^+$
production. Figs. 1(b) and 1(c) are, respectively, for the
$\tau^+$ and $e^+$ productions. Figs. 1(d), {\it etc.,} are of higher
orders in the perturbation language. We caution that evaluation of
these diagrams should begin with that the neutrino flavor states
are linear combinations of the neutrino mass eigen-states, with the
latter the "eigen-states" as in Dirac equations.

\begin{figure}[h]
\centering
\includegraphics[width=5in]{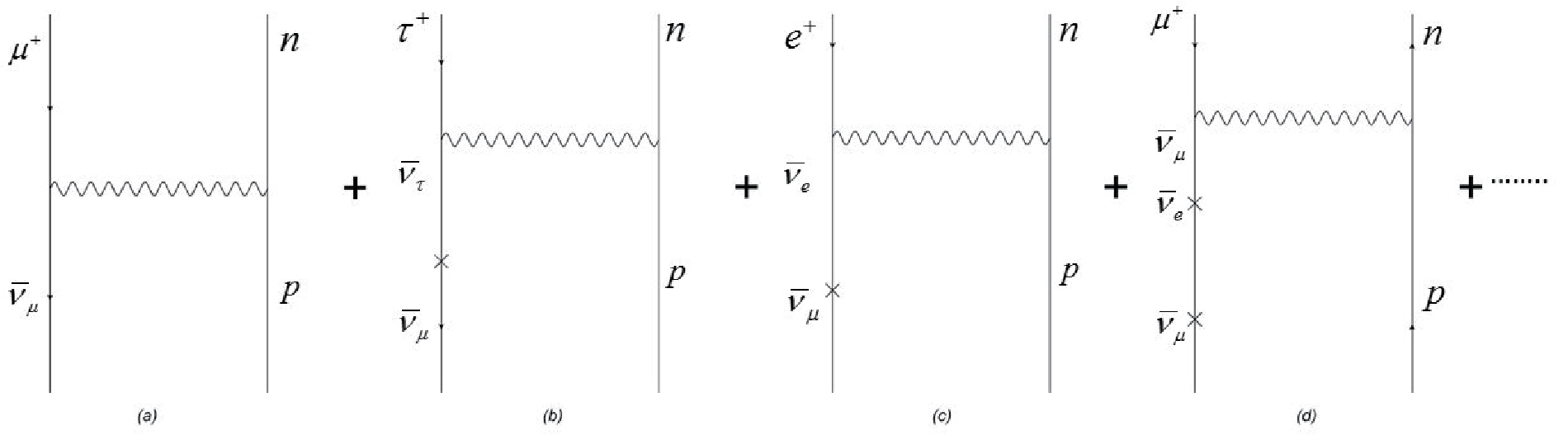}
\caption{The diagrams for the ${\bar \nu}_\mu p$
scattering.}
\end{figure}

Anyway we use the language in the diagonal case, to get some feeling.
To describe the reaction ${\bar\nu}_\mu (k) + p(p) \to \mu(k')
+ n (p')$ as shown in Fig. 1(a), we introduce \cite{Primakoff}

\begin{eqnarray}
&&<n(p')\mid I_\lambda^{(-)}(0)\mid p(p)>\nonumber\\
=&& i {\bar u}(p') \{ \gamma_\lambda f_V(q^2) - {\sigma_{\lambda\eta}
q_\eta \over 2 m_p} f_M(q^2)\} u(p),\\
&& <n(p')\mid I_\lambda^{5(-)}(0)\mid p(p)>\nonumber\\
= && i {\bar u}(p') \{ \gamma_\lambda \gamma_5 f_A(q^2) + {i2M
q_\lambda \gamma_5\over m_\pi^2} f_P(q^2) \}u(p),
\end{eqnarray}
with $q_\lambda=(p'-p)_\lambda$ and $2M=m_p+m_n$. Here $f_V(q^2)$,
$f_M(q^2)$, $f_A(q^2)$, and $f_P(q^2)$ are, respectively, the vector,
weak magnetism, axial, pseudoscalar form factors \cite{Primakoff}. The
transition amplitude is given by
\begin{eqnarray}
&&T({\bar\nu}_\mu p \to \mu^+ n)\nonumber\\
=&& {G_F cos\theta_c\over \sqrt 2} i \cdot "{\bar v}_\nu(k)" \cdot \gamma_\lambda
(1+\gamma_5) v_\mu(k')<n(p')\mid \{I_\lambda^{(-)}(0) + I_\lambda^{5(-)}(0)
\}\mid p(p)>.
\end{eqnarray}
Here, in the presence of neutrino oscillations, $"{\bar v}_\nu(k)"$
is {\it not} defined in our textbooks and should be replaced by
the linear combination $U_{\mu i}{\bar v}^i(k)$ with the neutrino
mass eigen-states $v^i(k)$. In other words, only the mass eigen-states
are introduced, not the flavor states, in our textbooks.

Thus, we obtain the following result \cite{Hwang9}.
\begin{eqnarray}
&& {d\sigma\over d\Omega_\mu}(\nu_\mu + p \to \mu^+ + n)\nonumber\\
= && {G_F^2cos^2\theta_c E^2_\mu\over 2\pi^2} {E_\mu\over E_\nu}\{
[f_V^2 + f_M^2 {q^2\over 2m_p} + f_A^2+ {m_\mu^2\over m_\pi^2}
(f_P^2 {q^2\over m_\pi^2}- 2f_Af_P)] cos^2 \theta_\mu\nonumber\\
&&+ 2 [(f_V+f_M)^2{q^2\over 4 m_p^2} + f_A^2(1+ {q^2\over 4m_p^2})
-4{E_\mu\over m_p}(1+{E_\nu\over m_p}sin^2{\theta_\mu\over 2})f_A(f_V+f_M)]
 sin^2{\theta_\mu\over 2}\}\nonumber\\
\equiv && {G_F^2 cos^2\theta_c E^2_\mu \over 2\pi^2} {E_\mu\over E_\nu}
\cdot N_0 \nonumber\\
= && (1.857\times 10^{-38} cm^2)\cdot N_0; \quad (E_\mu,\,E_\nu\,\, in\,\, 3\, GeV).
\end{eqnarray}

In the present case, we could treat the variables of the neutrino
as "dummy", that is, expressing them in terms of the other three
particles, $p$, $n$, and $\mu$. Thus, we could sidetrack the
phenomenon of neutrino oscillations for the problems of this kind.
Note that we cannot prepare the $\nu_\mu$ beam either, in
view of neutrino oscillations; so, the above
calculations do not have a direct meaning.

In fact, this suggest how to get around when the neutrino
variables could be treated as dummy variables. The present
example is a perfect example for that, but we are
calculating something not directly for the experiments.
However, if the neutrino is the only particle in the
process that does not have a well-defined mass, is the
sacred energy-momentum conservation law violated?

On the other hand, we proceed to consider the muon-like
neutrino converting to a tau-like neutrino and
then producing a tau, i.e. $({\bar\nu}_\mu \to {\bar\nu_\tau}) + p \to \tau^+
+ n$, as shown by Fig. 1(b). Feynman's rules yield
\begin{eqnarray}
iT=&& (-) {G_F\over \sqrt 2}\cdot {h\over 2}{1\over \sqrt 2} u_3\cdot
\sum_{i,j}{\bar v}_i(k) U^\dagger_{\mu i}\cdot
{1\over 2}(1-\gamma_5) \cdot
U_{\tau j}{1\over i} {m_j-i \gamma\cdot k\over m_j^2+k^2 -i\epsilon}\cdot \nonumber\\
&&\gamma_\lambda(1+ \gamma_5) v_\tau (k')\cdot <n(p')\mid [I_\lambda^{(-)}(0) +
I_\lambda^{5(-)}(0)]\mid p(p)>\nonumber\\
\sim && \{ {h\over 2}{1\over \sqrt 2} u_3 \}
\cdot {G_F \over \sqrt 2}\cdot \sum_{i,j}{\bar v}_i(k) U^\dagger_{\mu i}
U_{\tau j}{m_j \over m_j^2 +k^2 -i \epsilon}
\gamma_\lambda (1+ \gamma_5) v_\tau (k')\cdot
\nonumber\\
&&<n(p')\mid [I_\lambda^{(-)}(0) + I_\lambda^{5(-)}(0)] \mid p(p)>.
\end{eqnarray}
Here, for an estimate, the coupling $h$ is to be $\sim 0.1$ and
the condensate is approximately $\sim 200 \, GeV$ \cite{Origin}.
Taking $k^2$ the incoming momentum squared, the relative size
would be $h\cdot {u_3 m_j \over Q^2}$ on the amplitude.

It is clear that we could anticipate similar results for Fig. 1(c).
Thus, we may be able to handle Figs. 1(a)-1(c) altogether so that,
in correspondence, the "oscillating" beams could be handled
"altogether".

With the unitary transformation $U$ from the flavor states
to the mass eigen-states, the story might be very interesting
in understanding real quantum mechanics behaviors. Of course,
it would be extremely difficult to know all the matrix elements
$U_{a i}$, since it might be complex (in view of CP
violation and also of baryon asymmetry). Neutrino
oscillations would indeed offer us the playground
to know more in understanding the quantum phenomena.

Let's turn our attention to the other previously
well understood processes, such as the muon decay,
which becomes a little subtle \cite{Hwang417}.
For the basic processes such as the muon decay,
$\mu^- \to e^- + {\bar \nu}_e + \nu_\mu$, we may write,
symbolically, the transition amplitude \cite{Books}:
\begin{equation}
i T = {G_F \over \sqrt 2} {\bar u}_e(p') \gamma_\lambda(1+\gamma_5) v_e(k')
\cdot {\bar u}_{\nu_\mu} (k) \gamma_\lambda (1+\gamma_5) u_\mu(p) + others,
\end{equation}
but this is incorrect, since $u(p)$, $v(k)$, etc. are, by definition,
on the mass shells. Neutrino oscillations tell us $u_{\nu_\mu}(k) \equiv
U_{\mu i} u_i (k)$, with the left-hand side defined by "$\equiv$";
similarly for the antineutrino $v_e(k)$.

In fact, our language here is only for the mass-shell Dirac
spinors, not for something which oscillates. So, we should
write $\sum_i U_{e i} u^i(k)$ for the electron-like neutrino,
etc., since $u^i(k)$'s are the mass-eigenstates - that is how
we set up the Dirac equations for.

Let's remind us again that, for the muon or the electron,
they should be on mass shells in our language - that is
the way which we represent the muon or the electron.

Thus, we should write, for the muon decay,
\begin{eqnarray}
i T = & {G_F \over \sqrt 2} {\bar u}_e(p') \gamma_\lambda(1+\gamma_5) U_{e i} v_i(k')
\cdot {\bar u}_j (k) U^\dagger_{\mu j} \gamma_\lambda (1+\gamma_5) u_\mu(p) + \nonumber\\
  & {G' \over \sqrt 2} {\bar u}_e(p') (1-\gamma_5) U_{e i}v_i(k')
\cdot {\bar u}_j (k) U^\dagger_{\mu j} (1-\gamma_5) u_\mu(p).
\end{eqnarray}
Here the second term  is due to the family-Higgs exchange
$\eta'_1$. (Our $\eta'_1$ refers to the $\tau$ channel, by our
convention.) So, $G'$ [$\propto h^2/m_1^2$] is much smaller
than the Fermi coupling $G_F$.

The important point is that all the Dirac spinors are
on the mass shells, in writing the S-matrix elements
- so that the expressions can be further manipulated. 
In comparison, the flavor states could be written as the 
linear combinations, such as $\sum_i U_{\mu i} u^i(p,s)$ 
with $u^i(p,s)$ the Dirac spinors; and it seems no other 
way. So, we should not use "the flavor eigen-states" to
avoid confusion.

There is another important aspect related to ultra high energy
comic rays (UHECR's) - say, the proton of energy of
$10^{18}\,eV$ or higher. One might think that the proton is
stable, i.e., do not decay whatsoever. In the Universe, the
protons would encounter the matter medium and thus the
electrons - then $p(p_1) + e^- (p_2) \to n(p'_1) +
\nu_e(p'_2)$ in the extreme kinematics become possible.
The beam energy-momentum $({\vec P}, E)$ is so huge
compared to $M_W$. The coupling $g^2$, or $e^2$, does
not cut off (the strength) much.

So, we have
\begin{eqnarray}
{\vec P} &&= {\vec P}' + {\vec P}_\nu, \nonumber\\
\sqrt{m^2+P^2} + m_e && = \sqrt{m_n^2 +P'^2}+E_\nu.
\end{eqnarray}
Now, $E_\nu \approx  P_\nu$ because of the tiny neutrino
mass. One obtains
\begin{equation}
P_\nu = N/D, \qquad N=m^2-m_n^2+m_e^2+2m_e P(1+{m^2\over 2P^2}+...),
              \quad D= 2 m_e + ({m^2\over P}+ ...).
\end{equation}
Putting in the masses and the ultra high energy, say,
$P=10^{18}\,eV$, we obtain $P_\nu \approx P$. Thus,
we realize that the flow of the energy-momentum is
through the $W^+$ boson and then, almost completely,
into the neutrino. The final neutron, just like the
initial electron, serves as the spectator, no longer 
of initial ultra high energy.

The easy calculations then follow, for different
energies, etc. We could use the U-gauge for the leading
no-loop calculations - the $W$-mass parameter $M_W^2$
now is small in this region of the extreme kinematics. 
Our Universe is indeed rather interesting, when we 
consider the UHECR's behaviors of, e.g., $p+e^-\to 
n+\nu_e$.

{\it For the ultra high limits such as a proton of
$10^{18}\, eV$ or higher, the process $p+e^- \to
n+ \nu_e$ will help to deliver the proton energy
to the neutrino energy, in a weak process with the
energy so high that it is no longer
weak. The extreme kinematics, plus the lowest-order
$W$-graph, suggests that there may be important
mechanisms to cut off the ultra high energy limits.}

To close this section, we wish to reiterate \cite{Family1}
that neutrino oscillation is the "peculiar" change of
neutrino flavor, representing reshuffling of a combination
of several mass eigen-states to another combination. It
also implies a certain violation of lepton flavor for
these point-like Dirac particles, as explained in the 
next section.

\medskip

\section{Lepton-flavor violations}

Certain LFV processes \cite{PDG} such as $\pi^0 (\eta^0) \to \mu + e$,
as depicted by Fig. 2, $\mu \to e + \gamma$, as in Figs. 3, and $\mu
+ A \to A^* + e$, as shown in Figs. 4, are closely related to the most
cited picture of neutrino oscillations. The vertex for the
$\nu_\mu \to \nu_e$ transition is in fact coupled to some family Higgs
field with some VEV. Early on \cite{Family1,Hwang7}, it was realized that
the cross-generation or off-diagonal neutrino-Higgs interaction may
serve as the detailed mechanism of neutrino oscillations, with
some vacuum expectation value of the new Higgs field(s), granting that
neutrino are described as point-like Dirac particles.

In Fig. 2, we study the decay $\pi^0 (\eta^0) \to \mu + e$,
allowed if the family gauge theory exists in the lepton
world. In addition to experiments with neutrino oscillations,
such decays or reactions would be the first place to
look after.

\begin{figure}[h]
\centering
\includegraphics[width=1.5in]{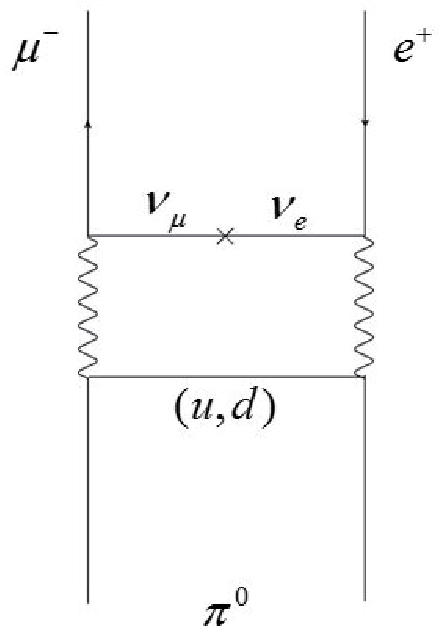}
\caption{The diagram for $\pi^0 (\eta^0) \to \mu + e$, showing a
close connection with neutrino oscillations.}
\end{figure}

For Fig. 2, we may write down the transition amplitude
via Feynman rules (Wu/Hwang \cite{Books}) as follows:
\begin{eqnarray*}
i T = && {1\over (2\pi)^4}\int d^4k \cdot {\bar u}(p) \cdot
i {1\over {2\sqrt 2}} {e\over sin \theta_W} \cdot i\gamma_\lambda (1+\gamma_5)\nonumber\\
&& \cdot \sum_{i,j} U^\dagger_{\mu i} {1\over i} {m_i-i\gamma\cdot k \over m_i^2+k^2-i\epsilon} \cdot i\cdot i{h\over 2}
{1\over \sqrt 2} {1\over 2}(1+\gamma_5) u_1 \nonumber\\
&& \cdot {1\over i} {m_j -i\gamma\cdot k \over m_j^2 +k^2 -i\epsilon} U_{e j}\cdot
i {1\over 2\sqrt 2}{e\over sin\theta_W}\cdot i\gamma_\eta(1+\gamma_5)\cdot v(p')\nonumber\\
&& \cdot {\bar v}(q') \cdot i {1\over 2\sqrt 2}{e\over sin\theta_W}\cdot i \gamma_\eta
(1+\gamma_5)\cdot {1\over i}{m-i\gamma\cdot (q+k-p) \over m^2+(q+k-p)^2-i\epsilon}\nonumber\\
&&\cdot i {1\over 2\sqrt 2}{e\over sin\theta_W}\cdot i \gamma_\lambda (1+\gamma_5)\cdot
u(q)\nonumber\\
&& \cdot {1\over i}{1\over M_W^2+(k-p)^2-i\epsilon}\cdot
{1\over i}{1\over M_W^2+(k+p')^2-i\epsilon}.
\end{eqnarray*}

The expression may be simplified as follows:
\begin{eqnarray}
i T = && ({1\over 2\sqrt 2}{e\over sin\theta_W})^4 \sum_{i,j} U^\dagger_{\mu i}
U_{e j} 4 i{h\over 2}{1\over \sqrt 2}
u_1\cdot m_i\nonumber\\
&& \cdot{1\over (2\pi)^4}\int d^4k {\bar u}(p)
\gamma_\lambda \gamma\cdot k \gamma_\eta(1+\gamma_5)\cdot v(p')\nonumber\\
&& \cdot {\bar v}(q') \gamma_\eta \gamma\cdot (q+k-p) \gamma_\lambda (1+\gamma_5)
u(q)\nonumber\\
&& \cdot {1\over m_i^2+k^2-i\epsilon}\cdot{1\over m_j^2+k^2-i\epsilon}
\cdot {1\over m^2+(q+k-p)^2 -i\epsilon}\nonumber\\
&& \cdot {1\over M_W^2+(k-p)^2-i\epsilon}\cdot
{1\over M_W^2+(k+p')^2-i\epsilon}.
\end{eqnarray}
It would be easy to guess the answer to be $G_F^2 {u_1 m_{i,j}/Q^2}$
(with $Q^2$ typical internal momentum transfer squared)
multiplied by a number of order unity. According to our work on the
origin of mass \cite{Origin}, the big numbers would be $M_W$ and
the condensate $u_1$; so, it is easy to guess the integral without
doing it explicitly.

The famous decay $\mu \to e + \gamma$ was investigated
by, e.g., \cite{HwangYan,Hwang7}. Here it is important
to note that the QED gauge invariance would make the
cancelation among different diagrams almost complete.
Thus the predicted decay rate would be uninteresting
small, experimentally.

In Figs. 3, we proceed to consider the golden lepton-flavor-violating
decay $\mu \to e + \gamma$, which were considered previously in
\cite{Hwang7} as well as in \cite{HwangYan}. The missing diagrams
are those with the W-boson exchanges replaced by the exchanges of
the charged family Higgs, again cutoff by gauge invariance - so, we
shall not consider the evaluation of these diagrams any further.

We show in Figures 3(a), 3(b), and 3(c) three leading basic "Feynman" diagrams.
Here the conversion of $\nu_\mu$ into $\nu_e$ is marked by a cross sign and it
is a term from our off-diagonal interaction given above with the family Higgs
vacuum expectation value $u_1$.

\begin{figure}[h]
\centering
\includegraphics[width=4in]{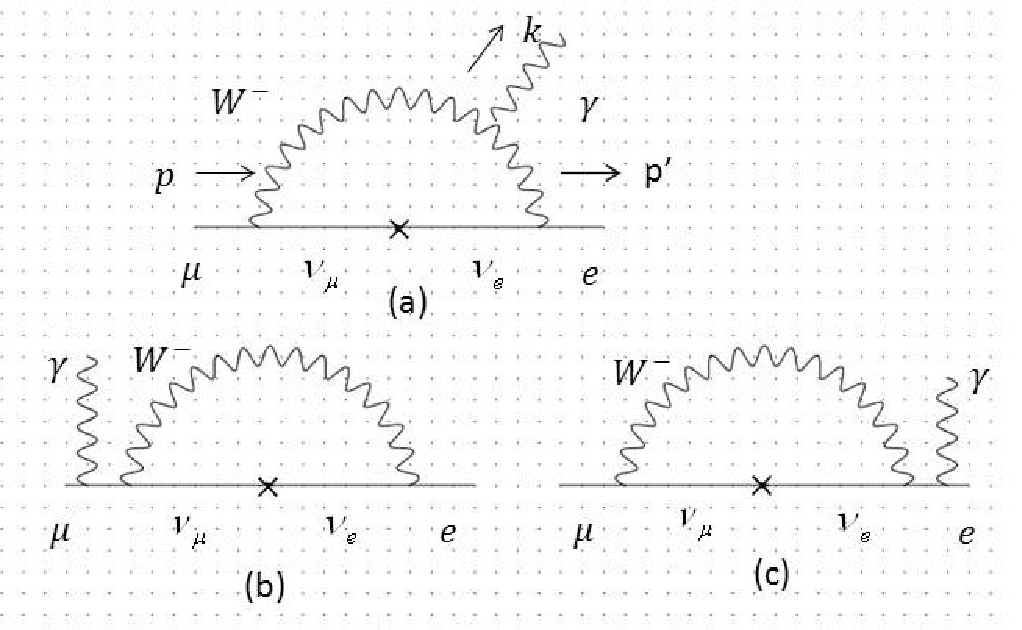}
\caption{The leading diagrams for $\mu \to e + \gamma$, noting
the cancelation among them due to QED gauge invariance for real
photons.}
\end{figure}

Using Feynman rules, we write, for Fig. 3(a),

\begin{eqnarray*}
{1\over (2\pi)^4} \int d^4q \cdot {\bar u}(p',s')\cdot &&i \cdot {1\over 2 \sqrt 2}
{e\over sin \theta_W}\cdot i \gamma_\lambda (1+ \gamma_5)\nonumber\\
\cdot \sum_{i,j} U^\dagger_{\mu i} {1\over i} {m_i-i\gamma\cdot q\over {m_i^2+q^2-i\epsilon}}\cdot
&& i \cdot i {h\over 2} {1\over \sqrt 2} {1\over 2} (1+\gamma_5)
\cdot U_{e j} {1\over i} {m_j-i\gamma\cdot q\over
{m_j^2 + q^2-i\epsilon}} \nonumber\\
\cdot i\cdot {1\over 2 \sqrt 2}{e\over sin\theta_W}\cdot &&i \gamma_{\lambda'}
(1+\gamma_5)\cdot u(p,s)\nonumber\\
\cdot {1\over i} {\delta_{\lambda'\mu}\over {M_W^2+(p-q)^2-i\epsilon}}\cdot
{\epsilon_\sigma(k)\over \sqrt{2k_0}}\cdot &&\Delta_{\sigma\mu\nu} \cdot
{1\over i} {\delta_{\nu\lambda}\over {M_W^2+(p-q-k)^2-i\epsilon}},
\end{eqnarray*}
with $\Delta_{\sigma\mu\nu}=(-ie)\{\delta_{\mu\nu}(-k-p-q)_\sigma +
\delta_{\nu\sigma}(p-q+p-q-k)_\mu +\delta_{\sigma\mu} (-p+q+k+k)_\nu \}$.

On the other hand, Feynman rules yield, for Fig. 1(b),
\begin{eqnarray*}
{1\over (2\pi)^4} \int d^4q \cdot {\bar u}(p',s')\cdot &&i \cdot {1\over 2 \sqrt 2}
{e\over sin \theta_W}\cdot i \gamma_\lambda (1+ \gamma_5)\nonumber\\
\cdot \sum_{i,j} U^\dagger_{\mu i}{1\over i} {m_i-i\gamma\cdot q\over {m_i^2+q^2-i\epsilon}}\cdot
&&i\cdot i {h\over 2} {1\over \sqrt 2} {1\over 2} (1+\gamma_5)
\cdot U_{e j}{1\over i} {m_j-i\gamma\cdot q\over
{m_j^2 + q^2-i\epsilon}} \nonumber\\
\cdot i\cdot {1\over 2 \sqrt 2}{e\over sin\theta_W}\cdot &&i \gamma_{\lambda'}
(1+\gamma_5)\cdot \nonumber\\
\cdot {1\over i} {\delta_{\lambda\lambda'}\over {M_W^2+(p'-q)^2-i\epsilon}}
\cdot {1\over i} {m_\mu - i\gamma\cdot p'\over {m_\mu^2+ p^{\prime 2}-i\epsilon}}
\cdot &&i (-i)e \gamma_\sigma\cdot {\epsilon(k)\over \sqrt {2k_0}}. u(p,s),
\end{eqnarray*}
and a similar result for Fig. 1(c).

The four-dimensional integrations can be carried out, via the dimensional
integration formulae (e.g. Ch. 10, Wu/Hwang \cite{Books}), especially
if we drop the small masses compared to the W-boson mass $M_W$ in the
denominator. In this way, we obtain, ignoring the variation over $m_i$,
\begin{eqnarray}
i T_a={G_F\over \sqrt 2} &&\cdot {h\over 2}{1\over \sqrt 2}
\cdot m_i\cdot (-2i){e\over (4\pi)^2}\nonumber\\
&&\cdot {\bar u}(p',s') {\gamma\cdot \epsilon\over \sqrt {2k_0}}
(1+\gamma_5) u(p,s).
\end{eqnarray}

It is interesting to note that the wave-function renormalization,
as shown by Figs. 3(b) and 3(c), yields
\begin{eqnarray}
i T_{b+c}= {G_F\over \sqrt 2} &&\cdot {h\over 2}{1\over \sqrt 2}\cdot m_i
\cdot (+2i){e\over (4\pi)^2}
\cdot \{{p'^2\over m_\mu^2 + p'^2} + {p^2\over m_e^2 + p^2}\}\nonumber\\ &&\cdot
{\bar u}(p',s') {\gamma \cdot \epsilon \over \sqrt{2k_0}} (1+\gamma_5) u(p,s),
\end{eqnarray}
noting that $p^2=-m_\mu^2$ and $p'^2=-m_e^2$ would make the contribution
from Figs. 3(b) and 3(c) to be the same as, but opposite in sign, that
from Fig. 3(a).

In a normal treatment, one ignores the wave-function renormalization
diagrams 3(b) and 3(c) in the treatment of the decays $\mu \to e + \gamma$,
$\mu \to 3e$, and $\mu+ A \to e+ A^*$. In that case, the cancelation would
not be there for $\mu\to e + \gamma$. This cancelation due to QED gauge
invariance makes any observation of this decay mode virtually impossible.
But for Fig. 2 (for $\pi^0\to \mu + e$) and for the $\mu\to e$ conversion
processes the cancelation is no longer there - we should be looking for
the lepton-favor violation at the {\it right} place.

Comparing this to the dominant mode $\mu \to e {\bar \nu}_e \nu_\mu$
\cite{Books}, we would obtain the branching ratio.
The decay rate for $\mu \to e+ \gamma$ would be of the order $(m_{neutrino}
\cdot m_\mu)^2/M_W^4$, which is extremely small.

The $\mu\to e$ conversion in a muonic atom might be
most interesting, for both (1) that the figure similar
to Fig. 2 is present, as shown in Figs. 4(d) and 4(e),
and (2) that QED gauge invariance does not operate, as
shown by Figs. 4(a)-4(c) where a real photon in Figs. 2
is replaced by the proton.

\begin{figure}[h]
\centering
\includegraphics[width=4in]{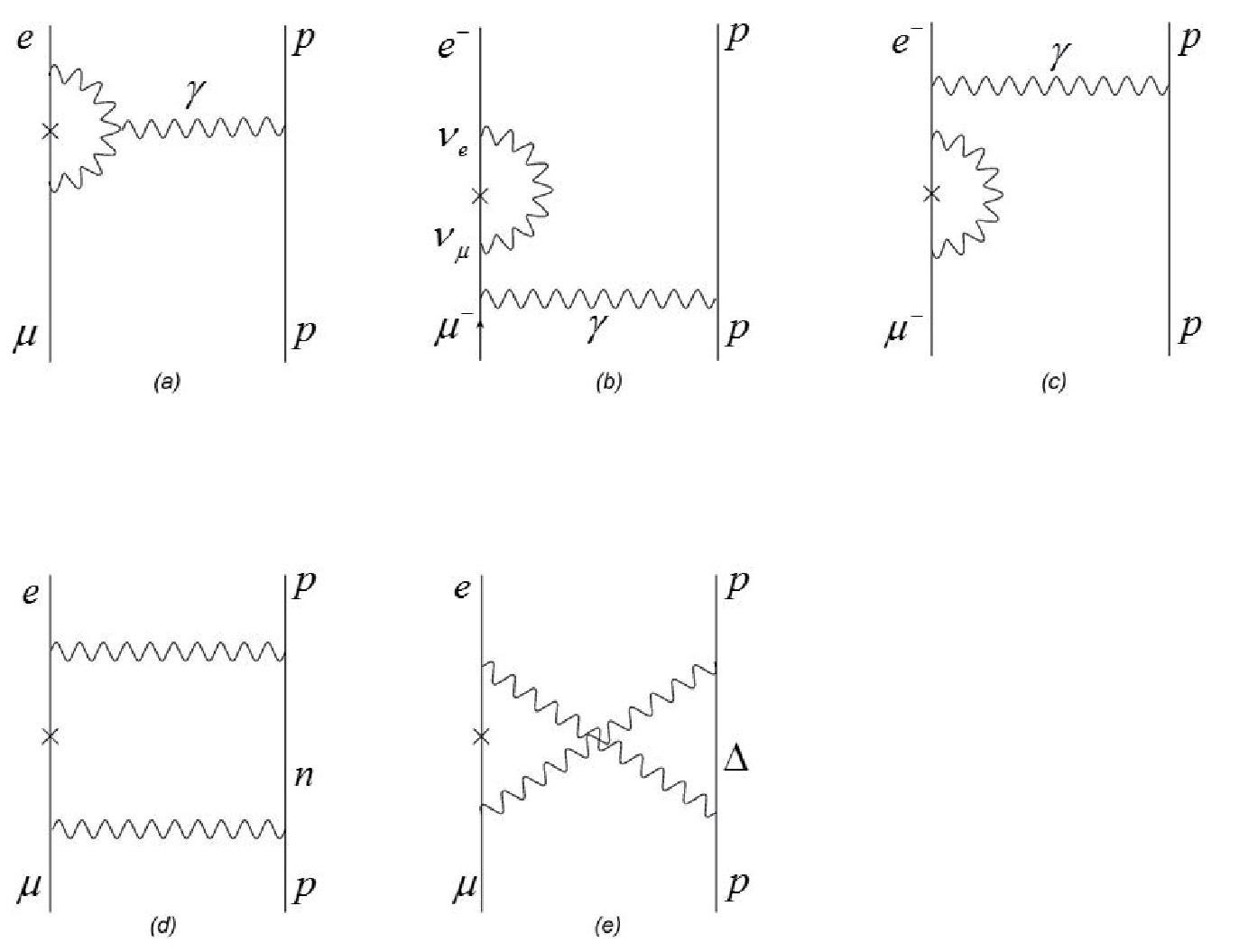}
\caption{The leading diagrams for $\mu + p \to e + p$, the
$(\mu \to e)$ conversion allowed if the family gauge symmetry
can be accommodated in the lepton world.}
\end{figure}

Thus, for Figs. 4(a)-4(c), the real photon in Figs. 3(a)-3(c) is replaced
by a virtual photon connected to the proton current (as in Figs. 1). The
Feynman rules can easily be modified. Physics-wise, the suppression
due to QED gauge invariance for real photons is no longer there.

On the other hand, for Fig. 4(d), it is the same mechanism as in
Fig. 2 (for the $\pi^0$ or $\eta^0$ decays). The result could be
duplicated here - with the result of order $G_F^2 (u_1 m_i/Q^2)$.
Fig. 4(e) is similar to Fig. 4(d) but it is smaller since it
involves the $\Delta^{++}$.

According to our estimate, the coupling $\kappa$ probably
brings down the cross sections by one or two orders of magnitudes;
as for powers of the Fermi coupling $G_F$ it could be the same order,
or slightly weaker, as the weak interactions. From analyzing Figs. 1-4,
the muon decay, etc., we feel rather optimistic in unraveling the
family gauge interactions, i.e., the effects
caused by family gauge bosons.

Besides the golden decays $\mu \to e+ \gamma$ (much too small, shown
in Figs. 3), $\pi^0 (\eta^0) \to \mu + e$ (as in Fig. 2), the
$\mu\to e$ conversion (as Figs. 4), and neutrino oscillations
(already observed, shown in Figs. 1), we also have to decide,
to what extent, violation of the $\tau-\mu-e$ universality. As
the baryon asymmetry is sometime attributed to the
lepton-antilepton asymmetry, the current scenario for neutrino
oscillations \cite{PDG} appears to be inadequate in this regard. If
we take the hints from neutrinos rather seriously, there are so
much to discover, even though the minimal Standard Model for the
ordinary-matter world remains to be pretty much intact.

\medskip

\section{Episode}

After the first version of this article appeared in 2012, there are four major
breakthroughs ("breakthroughs" in the eyes of the author) - first, Hwang
and Yan \cite{HwangYan} realized that three lepton $SU_L(2)$ doublets
could form an $SU_f(3)$ triplet and to make a consistent
theory one should start from the so-called "basic units" made up from
the right-handed Dirac components (of Dirac fields) or from the left-handed
Dirac components. And, secondly, the Higgs mechanisms are not independent
but, instead, are cooperative in the Standard-Model Higgs $\Phi(1,2)$,
purely family Higgs $\Phi(3,1)$, and the mixed Higgs $\Phi(3,2)$ in a
(new) Standard Model \cite{Hwang417}. Thirdly, we recognize that these
three Higgs could start from the massless phase and thus it explains
the origin of mass for everything \cite{Origin}. On the fourth, we
recognize that the 4-dimensional Minkowski space-time gives
the complex scalar fields a special status, in that the
self-repulsive interaction $\lambda (\phi^\dagger \phi)^2$ is
dimensionless and the value of $\lambda$ is given by the
4dimensional Minkowski space-time (and {\it not} by the field
itself). Thus, this is the origin of fields (point-like
particles) - all determined by the complex scalar Higgs fields
\cite{Fields}. In view of the self-repulsive nature of a Higgs
field, it require three Higgs fields $\Phi(1,2)$ (SM Higgs),
$\Phi(3,1)$ (purely family Higgs), and $\Phi(3,2)$ (mixed
family Higgs) to make up the story.

Our suggestion is as follows \cite{Fields}, \cite{Hwang417}:
We live in the quantum 4-dimensional
Minkowski space-time with the force-fields $SU_c(3) \times SU_L(2)
\times U(1) \times SU_f(3)$ gauge-group structure built-in from the
outset. The quark world and, separately, the lepton world are
admitted by this background of our world.

Under the assumption that, in the dark-matter world, the dark-matter
particles are also species in the extended Standard Model, most of
reactions happening among dark-matter particles, even involving
neutrinos, cannot be detected directly, due to the slowness, in the
ordinary-matter world. After all these developments, we could even
argue that the "minimum" extended Standard Model would be the extended
Standard Model to be based on the group $SU_c(3) \times SU_L(2)
\times U(1) \times SU_f(3)$ \cite{Hwang417}, provided that why
there are three generations of fermions could be explained. This
"minimal extended" Standard Model would be the most natural choice of
all the models.

My long journey with the family concept \cite{Family} finally ends up
with something specific, with understanding of the origin of mass
\cite{Origin}. Now, we turn to a precise definition of the Standard
Model - maybe a natural statement of the Standard Model could allow
us to understand why the 4-dimensional Mikowski space-time accepts
the Standard Model, but not something else. Thus, why our Space, i.e.,
the $SU_c(3) \times SU_L(2) \times U(1) \times SU_f(3)$, admits, and
only admits, these particles becomes the question that we could try to
ask and to answer. In this search, the journey becomes
awfully interesting in the long run.

How about the right-handed options \cite{Salam}? If we adopt the "basic units"
as Hwang and Yan \cite{HwangYan} proposed, the models would not belong since each
basic unit is associated with one kinetic term, one-to-one. Of course, the
projection operators could be deviated from the "right-handed" or "left-handed"
in the starting point; or, we do not start from those basic units. In any case,
the door seems to be shut up but, after different thoughts, there is another
door.

In a slightly different context \cite{Hwang3}, I proposed, five years
ago (due to the ignorance), to work with two working rules: "Dirac
similarity principle", based on eighty years of our experiences, and
"minimum Higgs hypothesis", from the last forty years of our another
experiences. By "Dirac similarity principle", all the fermions of the
extended Standard Model are point-like Dirac particles and this has
important implication for neutrinos. Using these two working rules, the
extended model mentioned above becomes rather unique - so, it is so
much easier to check it against the experiments. My long, and lonely,
journey became a guided tour.

It turns out that these two working rules, "Dirac similarity principle"
and "minimum Higgs hypothesis", coming from our experiences of eighty
years and of forty years, stands till today. So, neutrinos are
point-like Dirac particles. The complex scalar (Higgs) fields,
which cannot exist if alone, indeed claim the minimum existence.

We would be curious about how the dark-matter world looks like, though
it is difficult to verify experimentally. The first question would be if the
dark-matter world, 25 \% of the current Universe (in comparison, only 5 \%
in the ordinary matter), would clusterize to form the dark-matter
galaxies, maybe even before the ordinary-matter galaxies. The dark-matter
galaxies would then play the hosts of (visible) ordinary-matter galaxies,
like our own galaxy, the Milky Way. Note that a dark-matter galaxy is
by our definition a galaxy that does not "feel" any ordinary strong and
electromagnetic interactions (- unlike in our visible ordinary-matter
world). This fundamental quest deserves some thoughts, for the structural
formation of our Universe.

Of course, we should remind ourselves that, in our ordinary-matter world,
those quarks can aggregate in no time, to hadrons, including nuclei, and
the electrons serve to neutralize the charges also in no time; then atoms,
molecules, complex molecules, and so on. The early stage of "aggregation"
does not involve the gravitational force, and so much faster than those by
the gravitational force. This early stage of aggregation serves as the
"seeds" for the clusters, and then stars, and then galaxies, maybe in a time
span of $1\, Gyr$ (i.e., the age of our young Universe). The aggregation
caused by strong and electromagnetic forces is fast enough to help giving
rise to galaxies in a time span of $1\, Gyr$. On the other hand, the seeded
clusterings might proceed with abundance of extra-heavy dark-matter
particles such as familons and elusive family Higgs, all in the range
of a hundred $GeV$ or lighter if our work to understand the origin of
mass \cite{Origin} makes some sense, and with relatively long lifetimes
(owing to very limited decay channels). Thus, further simulations on
galactic formation and evolution may yield clues on our problem.

Finally, coming back to the fronts of particle physics, neutrinos,
especially the oscillating neutrinos, might couple to the other
particles in the dark-matter world. Neutrino oscillations and the family
"symmetry" were hints for us to enter the dark-matter world. Any further
investigation along this direction would be of utmost importance.

\bigskip

This research was supported in part by National Science Council project (NSC
99-2112-M-002-009-MY3). We wish to thank the authors of the following
books \cite{Books} for thorough write-ups on the minimal Standard Model.

\end{document}